\documentclass[preprint,prb,aps,floats,amssymb,showpacs,draft,%
floatfix,superscriptaddress]{revtex4} 

\begin{document}

\title{
22nd order high-temperature expansion of nearest-neighbor
models with O(2) symmetry on a simple cubic lattice.  
}

\author{Massimo Campostrini}
    \affiliation{Dipartimento di Fisica dell'Universit\`a di Pisa
        and I.N.F.N., I-56127 Pisa, Italy}
\author{Martin Hasenbusch}
    \affiliation{Dipartimento di Fisica dell'Universit\`a di Pisa
        and I.N.F.N., I-56127 Pisa, Italy}
\author{Andrea Pelissetto}
    \affiliation{Dipartimento di Fisica dell'Universit\`a di Roma I
        and I.N.F.N., I-00185 Roma, Italy}
\author{Ettore Vicari}
    \affiliation{Dipartimento di Fisica dell'Universit\`a di Pisa
        and I.N.F.N., I-56127 Pisa, Italy}

\begin{abstract}
  We present the high-temperature series for a nearest-neighbor model with O(2)
  symmetry on a simple cubic lattice with the most general single-site
  potential.  In particular, the magnetic susceptibility and the second-moment
  correlation length are computed to 22nd order.  The series specialized to
  some particular improved Hamiltonians have been already analyzed in the
  paper M. Campostrini, M.  Hasenbusch, A. Pelissetto, and E. Vicari, Phys.
  Rev. B 74, 144506 (2006) [cond-mat/0605083], to determine the critical
  exponents and other universal quantities of the three-dimensional XY
  universality class.
\end{abstract}

\maketitle

\vskip 1truecm

We have computed the high-temperature series for O(2)-symmetric
nearest-neighbor models on a simple cubic lattice with the most general
single-site potential.
We consider two-dimensional real vectors $\vec{\phi}_x$ defined at the sites 
$x$ of the lattice, the nearest-neighbor Hamiltonian 
\begin{equation}
{\cal H} =
 - \beta\sum_{\left<xy\right>} {\vec\phi}_x\cdot{\vec\phi}_y 
\end{equation}
and the partition function 
\begin{equation}
Z = \int \prod_x d\mu(\phi_x)\, e^{-{\cal H}}.
\end{equation}
Different models correspond to different choices of the single-site measure
$d\mu(\phi_x)$.
Three different models have been considered in the paper \onlinecite{CHPV-06}:
\begin{itemize}
\item[(i)] XY model:
\begin{equation}
d\mu(\phi_x) = d \phi_x^{(1)} \, d \phi_x^{(2)} \,
  \delta(1-|{\phi}_x|)\; ;
\end{equation}
\item[(ii)] ddXY model:
\begin{equation}
d\mu(\phi_x) = d \phi_x^{(1)} \, \phi_x^{(2)} \,
e^{D |{\phi}_x|^2}
\left[
\delta(\phi_x^{(1)}) \, \delta(\phi_x^{(2)})
 + \frac{1}{2 \pi} \, \delta(1-|{\phi}_x|)
\right];
\end{equation}
\item[(iii)] $\phi^4$ lattice model:
\begin{equation}
d\mu(\phi_x) = d \phi_x^{(1)} \, d \phi_x^{(2)} \,
   \exp{ \left[ - {\vec\phi}_x^2 - \lambda ({\vec\phi}_x^2 - 1)^2\right]}.
\end{equation}
\end{itemize}

Using the linked-cluster expansion technique, we have computed the
high-temperature expansion of several quantities.  
We considered the magnetic susceptibility $\chi$
\begin{equation}
\chi = \sum_x \langle \phi_{\alpha}(0) \phi_{\alpha}(x) \rangle,
\label{chi}
\end{equation}
and the zero-momentum connected $2j$-point Green's functions $\chi_{2j}$
($\chi = \chi_2$) for $j = 2,3,4,5$:
\begin{equation}
\chi_{2j} = \sum_{x_2,...,x_{2j}}
    \langle \phi_{\alpha_1}(0) \phi_{\alpha_1}(x_2) ...
        \phi_{\alpha_j}(x_{2j-1}) \phi_{\alpha_j}(x_{2j})\rangle_c.
\end{equation}
We also considered the first moments of the two-point function
\begin{equation}
m_{2k} = \sum_x x^{2k} \langle \phi_{\alpha}(0) \phi_{\alpha}(x) \rangle.
\end{equation}
More precisely, we computed $\chi$ to 22nd order, $\chi_4$ to 20th order,
$\chi_6$ and $\chi_8$ to 18th order, $\chi_{10}$ to 15th order, $m_2$ to 22nd
order, $m_4$ to 20th order, $m_6$ and $m_8$ to 19th order.

The high-temperature series specialized to some particular cases, providing
Hamiltonians with suppressed scaling corrections, have been already analyzed
in Refs.~\onlinecite{CHPV-06,CHPRV-01}.

The quantities necessary to reconstruct the series are reported in the TXT
files {\tt c\_phi4.TXT} and {\tt series.TXT} which can be found in
Ref.~\onlinecite{epaps}.

For each quantity the reported high-temperature series is written as 
\begin{equation}
O = \sum_{k,i_1,\ldots,i_{13}} \beta^k c_1^{i_1} \ldots c_{13}^{i_{13}}
      O(k;i_1,\ldots,i_{13})\; ,
\label{series}
\end{equation}
with coefficients $c_k$ defined below.
The presence of $c_k$ only for $k \le 13$ is of course related to the 
length of the present series: longer series require additional 
coefficients $c_k$. Indeed, in order to compute the $n$-th term in the 
expansion of $\chi_l$, one needs $c_k$ with $k \le \lfloor (n+l)/2\rfloor$.
For $m_{l}$, we have analogously $k \le \lfloor (n+1)/2\rfloor$,
independent of $l$.

The coefficients $c_k$ depend on the single-site measure and are defined as 
\begin{equation}
c_k = {1\over 2^k k!} {I_{1 + 2k}\over I_1} \qquad\qquad
I_k = \int_0^\infty x^k f(x),
\end{equation}
where $f(x)$ is defined by the single-site measure:
\begin{equation}
d\mu(\phi_x) = d \phi_x^{(1)} \, d \phi_x^{(2)} \,
   f(|\phi_x|).
\end{equation}
In file {\tt c\_phi4.TXT} of Ref.~\onlinecite{epaps} we report the quantities
$c_{k}$ for $k = 1,13$ for the $\phi_4$ theory with $1.90 \le \lambda \le
2.20$. Each line contains three numbers: $\lambda$, $k$, $c_k$.  For the ddXY
model, the coefficients $c_k$ can be computed analytically:
\begin{equation}
c_k = {1\over 2^k k!}  
   \left[ \delta_{k0} + (1 - \delta_{k0}) {1\over 1 + e^{-D}}\right]\; .
\end{equation}
For the XY model
\begin{equation}
c_k = {1\over 2^k k!} \; .
\end{equation}
The high-temperature series are reported in file {\tt series.TXT}
of Ref.~\onlinecite{epaps}. 
Each line contains 16 numbers: the first number indicates 
which observable one is referring to: 1 refers to $\chi$, 2 to $\chi_4$,
3 to $\chi_6$, 4 to $\chi_8$, 5 to $\chi_{10}$, 6 to $m_2$, 
7 to $m_4$, 8 to $m_6$, 9 to $m_8$. The following 14 numbers are 
respectively $k$, $i_1,\ldots,i_{13}$ in Eq.~(\ref{series}).
The last  number is the coefficient $O(k;i_1,\ldots,i_{13})$.
We only report the nonvanishing coefficients.

\end{document}